\documentclass[prl,twocolumn,amsmath,floats,showpacs,superscriptaddress]{revtex4}

\usepackage{graphicx}
\usepackage{epsfig}
\usepackage{dcolumn}
\input{epsf} 
\begin{document}
\title{Highly conductive molecular junctions based on direct binding of
benzene to platinum electrodes}

\author{M. Kiguchi}
\thanks{Present address: Graduate School of Science, Hokkaido University,
Sapporo, Japan, and JST-PRESTO}
 \affiliation{Kamerlingh Onnes
Laboratory, Leiden University, Leiden, The Netherlands}

\author{O. Tal}
\affiliation{Kamerlingh Onnes Laboratory, Leiden University,
Leiden, The Netherlands}

\author{S. Wohlthat}
\affiliation{School of Chemistry, The University of Sydney,
Sydney, Australia} \affiliation{Institut f\"ur Theoretische
Festk\"orperphysik and DFG-Center for Functional Nanostructures,
Universit\"at Karlsruhe, Karlsruhe, Germany}

\author{F. Pauly}
\affiliation{Institut f\"ur Theoretische Festk\"orperphysik and
DFG-Center for Functional Nanostructures, Universit\"at Karlsruhe,
Karlsruhe, Germany}

\author{M. Krieger}
\thanks{Present address: Institute of Applied Physics, University of
Erlangen-N\"{u}rnberg, Erlangen, Germany}
\affiliation{Kamerlingh
Onnes Laboratory, Leiden University, Leiden, The Netherlands}

\author{D. Djukic}
\affiliation{Kamerlingh Onnes Laboratory, Leiden University,
Leiden, The Netherlands}

\author{J.C. Cuevas}
\affiliation{Departamento de F\'{\i}sica Te\'orica de la Materia
Condensada, Universidad Aut\'onoma de Madrid, Madrid,
Spain}\affiliation{Institut f\"ur Theoretische Festk\"orperphysik
and DFG-Center for Functional Nanostructures, Universit\"at
Karlsruhe, Karlsruhe, Germany}

\author{J.M. van Ruitenbeek}
\affiliation{Kamerlingh Onnes Laboratory, Leiden University,
Leiden, The Netherlands}

\begin{abstract}
Highly conductive molecular junctions were formed by direct
binding of benzene molecules between two Pt electrodes.
Measurements of conductance, isotopic shift in inelastic
spectroscopy and shot noise compared with calculations provide
indications for a stable molecular junction where the benzene
molecule is preserved intact and bonded to the Pt leads via carbon
atoms. The junction has a conductance comparable to that for
metallic atomic junctions (around 0.1-1 G$_0$), where the
conductance and the number of transmission channels are controlled
by the molecule's orientation at different inter-electrode
distances.
\end{abstract}

\date{\today}
 \maketitle


Connecting a molecule as a bridge between two conducting
electrodes is one of the fundamental challenges involved with the
study of electron transport through molecular
junctions~\cite{Tao2006}. The difficulty stems from a variety of
requirements which are sometimes contradictory: a good
molecule-electrode contact should be easy to achieve, for example,
by chemical binding but chemically insensitive to environmental
influences, mechanically stable but flexible enough to allow
molecular rearrangement, provide a good electronic coupling
between the molecule and the conducting electrodes but still
preserve to some degree the individual electronic properties of
the molecule. A common approach in fabrication of such molecular
junctions utilizes functional side groups attached to the main
molecule structure as anchoring "arms" that chemically bind to
metallic leads (e.g thiol~\cite{Reed1997},
amine~\cite{Venkataraman2006} and carboxylic~\cite{Chen2006}
groups). The thiol group became the most widely used anchoring
group~\cite{Reed1997,Ulrich2006,Li2006,Li2008}, since it binds
readily to gold electrodes with some flexibility for rearrangement
while providing a finite electronic coupling between the molecule
and the leads. However the use of thiols and other anchoring side
groups poses some inherent disadvantages: the measured conductance
of thiol based molecular junctions is distributed over a wide
range of values~\cite{Ulrich2006,Li2006,Li2008} which is mainly
ascribed to different binding configurations on Au
electrodes~\cite{Li2008,Fujihira2006,Muller2006}. Although other
anchoring groups may overcome this
drawback~\cite{Venkataraman2006}, in general the anchoring groups
act as resistive spacers between the electrodes and the molecule.
This leads to low conductivity and sensitivity to different
environmental effects such as neighbor adsorbed
species~\cite{Long2006}.

Here we report on a highly conductive molecular junction achieved
by direct binding of a $\pi$-conjugated organic molecule (benzene)
to metallic electrodes (Pt) without the use of anchoring groups.
The formation of such a molecular junction is verified by
conductance measurements as well as the effect of isotopic
substitution on inelastic electron spectroscopy. The evolution of
conductance, vibration modes and transmission channels as a
function of the inter-electrode distance is studied by comparing
measurements of conductance, inelastic electron spectroscopy and
shot noise to theoretical calculations.

\begin{figure}[t!]
\begin{center}
\includegraphics[width=6.0cm]{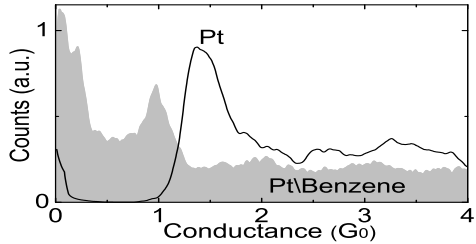}
\end{center}
\caption{Conductance histograms (normalized to the area under the
curves) for a Pt junction (black), and for Pt after introducing
benzene (filled). Each conductance histogram is constructed from
more than 3000 conductance traces recorded with a bias of 0.1 V
during repeated breaking of the contact.}\label{fig1}
\end{figure}

The measurements were performed using mechanically controllable
break junctions~\cite{Muller1992}. A notched Pt wire (0.1mm
diameter, 99.99\% purity) was glued on top of a bending beam and
mounted in a three-point bending configuration inside a vacuum
chamber. Once under vacuum and cooled to 4 K, the wire was broken
by mechanical bending of the substrate. The bending can be relaxed
to form an atomic-sized junction between the Pt wire ends using a
piezo element for fine adjustment. The formation of a clean Pt
junction is verified by conductance histograms made from more than
3000 conductance traces taken during repeated junction stretching,
as presented in Fig. \ref{fig1} (black curve). The single peak
around 1.4-1.5 G$_0$ (G$_0$=$2e^{2}/h$ is the conductance quantum)
is the typical signature of a clean Pt junction
~\cite{Djukic2005}.

Benzene (Sigma-Aldrich, purity$\geq$99.9$\%$) was degassed through
repeated freeze-pump-thaw cycles. Following the formation of the
Pt junction, the benzene was admitted using a leak valve via a
heated capillary to the Pt junction while the latter is broken and
formed repeatedly. During the benzene introduction, the typical Pt
peak is observed to be suppressed, and a single peak appears near
1 G$_0$ accompanied with a low conductance tail (Fig. \ref{fig1},
filled curve). In some cases, the histogram exhibits a peak near
0.2 G$_0$ on top of the tail. These findings imply that after the
introduction of benzene, the formation of pure Pt junctions is
suppressed while new junctions with preferred conductance of 1
$G_0$ and sometimes 0.2 $G_0$ are formed while stretching the
contact.

\begin{figure}[t!]
\begin{center}
\includegraphics[width=8.0cm]{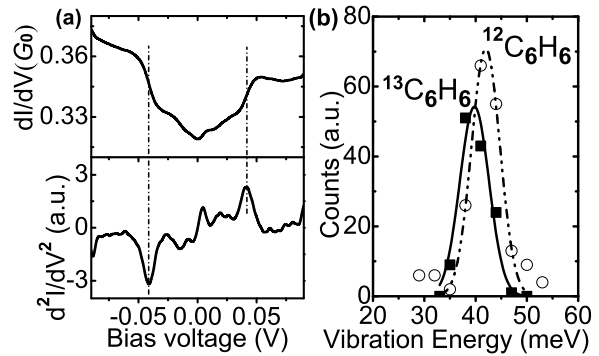}
\end{center}
\caption{(a) Differential conductance (top) and its derivative
(bottom) for Pt contact after introduction of benzene taken at a
zero bias conductance of 0.3 G$_0$. (b) Distribution of vibration
energy at low conductance regime.} \label{fig2}
\end{figure}

Differential conductance ($dI/dV$)~\cite{Smit2002} was measured as
a function of voltage across the Pt/benzene junctions where each
curve was recorded at fixed electrodes separation. As demonstrated
at the top of Fig. \ref{fig2}(a) a symmetric upward step in the
differential conductance is observed at 40~mV (as can be
determined more accurately by the corresponding peaks in the
derivative ($d^{2}I/dV^{2}$) in the lower panel of Fig.
\ref{fig2}(a)). The step feature indicates on a vibration which is
excited by the transport electrons having an energy of 40
meV~\cite{Djukic2005}. The conductance enhancement is typical for
inelastic electron tunneling spectroscopy, as expected for
vibration excitation at zero bias conductance below 0.5 G$_0$
~\cite{Paulsson2005,Tal2008CondMat}. In order to accurately
determine the associated vibration energy 250 differential
conductance spectra were collected for junctions having a zero
bias conductance of 0.05-0.4 G$_0$, resulting in the energy
histogram presented in Fig. \ref{fig2}(b) (circles). The histogram
shows a well-defined peak at 42 meV with 5 meV width. The
vibration energy for a given junction is insensitive to junction
stretching, unlike the vibration modes for Pt/H$_2$
junctions~\cite{Djukic2005}. We also measured the differential
conductance spectra for junctions with a conductance close to 1
G$_0$, but the positions in energy of the conductance steps do not
reveal any well-defined vibration energies, in contrast to the
junctions having a conductance below 0.4 G$_0$ (Fig.
\ref{fig2}(b)).

We repeated the experiment using $^{13}$C isotope substitution in
all six positions of benzene (Buchem, 99$\%$$^{13}$C$_{6}$H$_{6}$,
purity$\geq$98$\%$). The mass ratio of $^{13}$C$_{6}$H$_{6}$ and
$^{12}$C$_{6}$H$_{6}$ is 84/78, from which we predict a shift of
the vibration mode from 42 meV by the square root of this ratio to
40 meV for a vibration mode that involves the movement of the
whole molecule. The Pt/$^{13}$C$_{6}$H$_{6}$ energy histogram
obtained from 130 differential conductance spectra (filled squares
in Fig. \ref{fig2}(b)) shows a peak at 40 meV with 5 meV width, in
excellent agreement with the shift expected for a molecular
vibration mode which involves the movement of the entire benzene
molecule. This observation supports the picture of transport
electrons crossing a junction composed from a benzene molecule.

Focusing on the way that electrons are transmitted through the
junction we turn to shot noise measurements. As the shot noise
level is determined by the number of available transmission
channels across the junction and their transmission probabilities,
${\tau_{i}}$, the main transmission probabilities can be resolved
by comparing noise and conductance measurements to the expression
for noise power for an arbitrary set of transmission
probabilities, at temperature $T$ and bias voltage $V$ as given
by~\cite{Blanter2000}:
\begin{equation}S_I=2eV{\rm coth}\left(\frac{eV}{2kT}\right)\frac{2e^{2}}{h}
\sum_{i}\tau_{i}(1-\tau_{i})+4kT\frac{2e^{2}}{h}\sum_{i}\tau_{i}^{2}\end{equation}
where $k$ is Boltzmann's constant. Using the method described in
Refs.~\cite{Djukic2006,Tal2008CondMat} we have measured the shot
noise (i.e. $S_I$(V$>$0)$-$$S_I$(V$=$0)) across junctions with
different conductance values. Once a stable junction was
established at a certain conductance, the noise power was measured
as a function of frequency at different bias currents. At each
bias 10,000 noise spectra where averaged. Differential conductance
spectra were measured before and after every set of noise
measurements to verify that the same contact was maintained during
the measurements.

\begin{figure}[t!]
\begin{center}
\includegraphics[width=8.0cm]{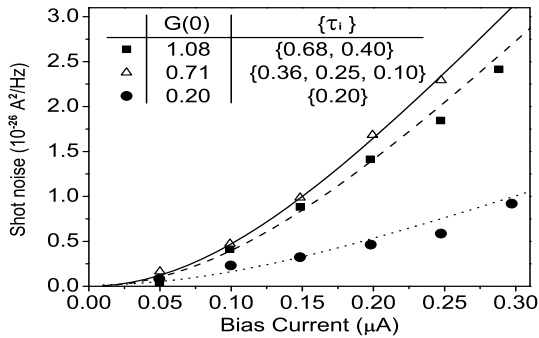}
\end{center}
\caption{Shot noise as a function of the bias current across the
Pt/benzene junction. The zero bias conductance (G(0)) for this
junction is 1.08 G$_0$ (filled squares), 0.71 G$_0$ (hollow
triangles) and 0.2 G$_0$ (filled circles). Fitting the data with
theory (curves) gives the decomposition of the total transmission
in terms of transmission probabilities, ${\tau_{i}}$, of the
conduction channels (where G$=$$\sum_{i}\tau_{i}$G$_0$) as shown
in the inset.} \label{fig3}
\end{figure}

Figure \ref{fig3} presents three sets of shot noise measurements
as a function of bias current. For junctions with zero bias
conductance of 1.08 (squares), 0.71 (open triangles), and
0.20$\pm$0.01 G$_0$ (bullets) the following transmission
probabilities were obtained : \{0.68, 0.40\}, \{0.36, 0.25, 0.10\}
and \{0.20\} respectively. The number of channels is eventually
reduced to one when the conductance is reduced to 0.2 G$_0$, while
at higher conductance (also well below 1 G$_0$) multiple channels
make up the transport across the junction. As opposed to Pt/H$_2$O
junctions there is no dominant transmission channel when more than
a single channel exists~\cite{Tal2008CondMat}. Note that the
uncertainty in the value of the main transmission probability is
small (in this case to $\pm$0.01), since the fit is very sensitive
to the number of channels and their
probabilities~\cite{vandenBrom1999,Djukic2006}. Choosing more
channels than presented in Fig. \ref{fig3} (inset) is restricted
to small additional channels that do not affect the specified
probabilities (within $\pm$0.01).

In order to better understand the experimental observations in
terms of a Pt/benzene junction we have performed extensive DFT
calculations using TURBOMOLE v5.7~\cite{Ahlrichs1989}, where we
have used a split valence polarization basis set for all
non-hydrogen atoms~\cite{Schaefer1992} and the BP86
exchange-correlation functional~\cite{Perdew1986}. To determine
the possible geometries of the molecular junction and its
evolution upon stretching we proceed as follows. We first place a
benzene molecule in between two Pt clusters composed by 8 atoms
sitting in the sites of a fcc lattice, which simulate two
atomically sharp electrode tips oriented in the (111) direction
with a lattice parameter of 3.92 {\AA}. Then, the geometry is
optimized holding the Pt electrodes fixed. Finally, we move apart
the pyramids stepwise and relax the molecule in each step.

\begin{figure}[t!]
\begin{center}
\includegraphics[width=8.0cm]{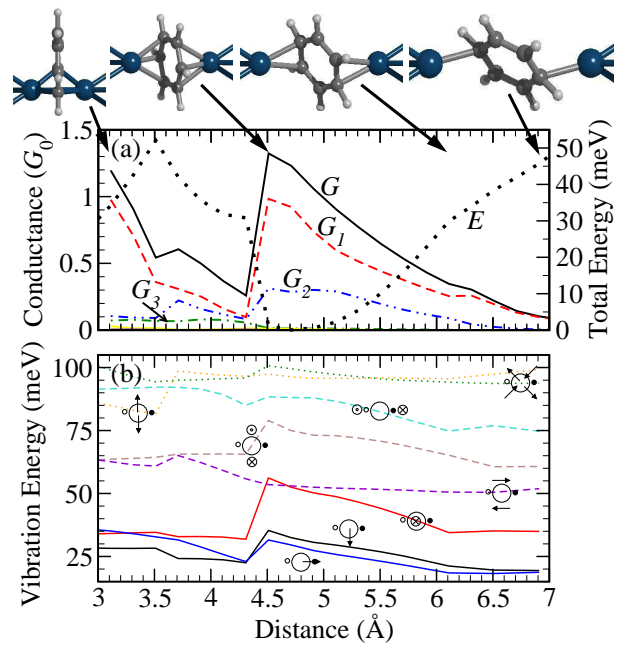}
\end{center}
\caption{(Color online) Simulation of the stretching process of a
Pt/benzene junction. (a) Total conductance, $G$, the contribution
of the individual conduction channels, G$_i$$=$G$_0\tau_i$, and
the change of the total energy, $E$, as a function of the distance
between the Pt tip atoms. (b) Stretching dependence of the energy
of the benzene vibration modes in the range between 0 and 105 meV.
The translation modes are plotted as solid lines, the rotation
modes as dashed lines and the stretch modes as dotted lines. The
character of the modes is indicated by symbols, where benzene is
represented by a large circle, the Pt atoms by two small circles
and arrows depicted the motion of the molecule.} \label{fig4}
\end{figure}

Although other junction geometries are possible, we focused on the
analysis of the junction evolution during stretching shown at the
top of Fig. \ref{fig4}, which seems to be compatible with many of
the experimental observations described above. Initially, the
benzene sits on top of the Pt electrodes with its plane
perpendicular to the Pt junction axis and bonded to both pyramids
with two C atoms. At an inter-electrodes distance of 4.5 {\AA} the
molecule jumps to a central position between the electrodes to
form a molecular junction. During the junction stretching the
molecule is progressively tilted while the number of C atoms that
bind to each Pt electrode~\cite{note1} is reduced from three to
two and eventually to one. Notice that during the junction
evolution the benzene retains its molecular integrity (in
particular, there is no dehydrogenation). The total energy has a
minimum  shortly after the formation of the benzene bridge
(Fig.~\ref{fig4}(a), dots), which indicates the most stable
configuration.

For each geometry we have computed the vibration modes of the
junction (Fig. \ref{fig4}(b)), keeping all Pt atoms fixed, which
is justified by the large difference in mass between C or H and
Pt. At electrode separations larger than 4.5 {\AA} there are two
possible candidates for the mode observed in the experiment. The
first one is a longitudinal mode, where the entire molecule
oscillates back and forth along the junction axis between the
electrodes. This mode becomes softer as the junction is elongated
leading to energy decrease from 50 meV to 35 meV at large
distances. The second mode involves a molecular rotation with
respect to its C$_6$-axis, having an energy around 50 meV. This
energy is rather insensitive to elongation of the contact,
involves the motion of the whole molecule, and has the appropriate
energy in the appropriate conductance range (see below). For these
reasons, this vibration is our main candidate for the mode
observed in the experiment.

To calculate the conductance of these molecular junctions we have
used Green's function techniques and the Landauer formula
expressed in a local non-orthogonal basis, as discussed in detail
in Refs.~[\onlinecite{Wohlthat2007, Pauly2007}]. We obtain the
bulk Green's functions following the procedure described in
Ref.~[\onlinecite{Pauly2007}] by computing separately the
electronic structure of a spherical Pt fcc cluster with 459 atoms.
Additionally, we obtain from this analysis a value for the Fermi
energy of -5.4 eV.

In Fig.~\ref{fig4}(a) we show the calculated results for the
conductance and its decomposition into conduction channels. In the
range where the benzene molecule sits on top of the Pt-Pt bond the
conductance decreases with stretching from a value above 1 G$_0$
down to 0.2 G$_0$, while the current is mainly carried by three
channels. When the benzene moves in between the electrodes to form
a molecular junction, the conductance jumps up to 1.3 G$_0$. Then,
when the total energy reaches its minimum (i.e. the most stable
geometry), the conductance is very close to 1 G$_0$, which may
explain the main peak in the conductance histogram (Fig.
\ref{fig1})). As the stretching proceeds, the conductance
decreases monotonically and the current is mainly controlled by
two channels and finally by only one in the last stages. We find
some similarity between the channel decomposition obtained by
noise measurements and the calculated one at inter-electrodes
distances of 3.4, 4.9, and 6.6 {\AA} for the junctions with
conductance of 0.71, 1.08 and 0.20 G$_0$, respectively. In
general, the calculated transmission probabilities indicate a lack
of a dominant channel when more than a single channel exist and a
reduction of the number of channels when the conductance of the
molecular junction (for distances larger than 4.5 {\AA}) is
reduced as was also implied by the experimental findings.
Interestingly, as a rule of thumb, an upper limit to the number of
conduction channels when the molecular junction is formed, is
simply given by the number of C atoms bonded to the Pt tip atoms.
This can be understood as follows. Since each C atom has only one
orbital taking part in the $\pi$-system of the benzene ring, each
atom can build at most one $\pi$-channel. For distances smaller
than 4.5 {\AA}, channels involving $\sigma$-bonds might play a
role, as well as direct tunneling from one Pt electrode to the
other.

Our experimental observations together with the agreement in many
points between the experimental results and the calculations
indicate the formation of a stable and highly conductive junction
by binding a benzene molecule to the platinum electrodes via Pt-C
bonds while preserving its molecular structure. Stretching of the
junction leads to tilting of the molecule which reduces both the
conductance and the number of transmission channels across the
junction as a consequence of sequential breaking of the Pt-C
bonds.

This work is part of the research program of the ``Stichting
FOM'', which is financially supported by NWO. OT is also grateful
for the AVS-Welch scholarship. MK greatly acknowledges the support
by the European Commission (RTN DIENOW). SW acknowledges the
Australian Partnership for Advanced Computing for computing
resources, FP the funding of a Young Investigator Group at KIT and
JCC the financial support of the EU network BIMORE
(MRTN-CT-2006-035859).


\begin{thebibliography}{24}
\bibitem[{not()}]{note1}
\bibinfo{note}{We consider that a C atom is coupled to the Pt electrodes if the
  Pt-C bond distance is smaller than 2.65 {\AA}.}
\end{thebibliography}
\end{document}